\documentclass[twocolumn,showpacs,amsmath,amssymb,pra,superscriptaddress]{revtex4-2}

\usepackage[T1]{fontenc}
\usepackage{lmodern}
\usepackage{amssymb}
\usepackage{amsmath}
\usepackage{textgreek}
\usepackage{microtype}
\usepackage {longtable}
\usepackage{bm}
\usepackage{float}
\usepackage{graphicx}
\usepackage{xcolor}
\usepackage{csquotes}
\usepackage[hyperindex,breaklinks]{hyperref}
\newcommand{\ket}[1]{|#1\rangle}

\newcommand \be{\begin{equation}}
\newcommand \ee{\end{equation}}
\newcommand \bea{\begin{eqnarray}}
\newcommand \eea{\end{eqnarray}}
\newcommand \bse{\begin{subequations}}
\newcommand \ese{\end{subequations}}

\begin{document}
\relpenalty=1
\title{Toffoli gate based on a three-body fine-structure-state-changing F\"{o}rster resonance in Rydberg atoms}

\author{I.~N.~Ashkarin}
\email{ivan.ashkarin@universite-paris-saclay.fr}
\affiliation {Rzhanov Institute of Semiconductor Physics SB RAS, 630090 Novosibirsk, Russia}
\affiliation {Novosibirsk State  University, Faculty of Physics, 630090 Novosibirsk, Russia}
\affiliation {Laboratoire Aim\'e Cotton, CNRS, Univ. Paris-Sud, ENS-Cachan, Universit\'e Paris-Saclay, 91405 Orsay, France}

\author{I.~I.~Beterov}
\affiliation {Rzhanov Institute of Semiconductor Physics SB RAS, 630090 Novosibirsk, Russia}
\affiliation {Novosibirsk State  University, Faculty of Physics, 630090 Novosibirsk, Russia}
\affiliation {Institute of Laser Physics SB RAS, 630090 Novosibirsk, Russia}
\affiliation {Novosibirsk State Technical University, Department of Physical Engineering, 630073 Novosibirsk, Russia}

\author{E.~A.~Yakshina}
\affiliation {Rzhanov Institute of Semiconductor Physics SB RAS, 630090 Novosibirsk, Russia}
\affiliation {Novosibirsk State  University, Faculty of Physics, 630090 Novosibirsk, Russia}
\affiliation {Institute of Laser Physics SB RAS, 630090 Novosibirsk, Russia}

\author{D.~B.~Tretyakov}
\affiliation {Rzhanov Institute of Semiconductor Physics SB RAS, 630090 Novosibirsk, Russia}
\affiliation {Novosibirsk State  University, Faculty of Physics, 630090 Novosibirsk, Russia}

\author{V.~M.~Entin}
\affiliation {Rzhanov Institute of Semiconductor Physics SB RAS, 630090 Novosibirsk, Russia}
\affiliation {Novosibirsk State  University, Faculty of Physics, 630090 Novosibirsk, Russia}

\author{I.~I.~Ryabtsev}
\affiliation {Rzhanov Institute of Semiconductor Physics SB RAS, 630090 Novosibirsk, Russia}
\affiliation {Novosibirsk State  University, Faculty of Physics, 630090 Novosibirsk, Russia}

\author{P.~Cheinet}
\affiliation {Laboratoire Aim\'e Cotton, CNRS, Univ. Paris-Sud, ENS-Cachan, Universit\'e Paris-Saclay, 91405 Orsay, France}

\author{K.-L.~Pham}
\affiliation {Laboratoire Aim\'e Cotton, CNRS, Univ. Paris-Sud, ENS-Cachan, Universit\'e Paris-Saclay, 91405 Orsay, France}

\author{S.~Lepoutre}
\affiliation {Laboratoire Aim\'e Cotton, CNRS, Univ. Paris-Sud, ENS-Cachan, Universit\'e Paris-Saclay, 91405 Orsay, France}

\author{P.~Pillet}
\affiliation {Laboratoire Aim\'e Cotton, CNRS, Univ. Paris-Sud, ENS-Cachan, Universit\'e Paris-Saclay, 91405 Orsay, France}

\begin{abstract}

We have developed an improved scheme of a three-qubit Toffoli gate based on fine structure state changing three-body Stark-tuned Rydberg interaction. This scheme is a substantial improvement of our previous proposal [I.I.Beterov et al., Physical Review A 98, 042704 (2018)]. Due to the use of a different type of three-body F\"{o}rster resonance we substantially simplified the scheme of laser excitation and phase dynamics of collective three-body states. This type of F\"{o}rster resonance exists only in systems with more than two atoms, while the two-body resonance is absent. We reduced the sensitivity of the gate fidelity to fluctuations of external electric field and eliminated the necessity to use external magnetic field for fine tuning of the resonant electric field value, compared to the previous scheme of Toffoli gate based on Rydberg atoms. A gate fidelity of~>99\% was demonstrated in the calculations.
 
\end{abstract}
\maketitle

%=========================================================================================================
%=========================================================================================================
%=========================================================================================================
\begin{center}
\begin{figure*}[!ht]
\center
\includegraphics[width=\textwidth]{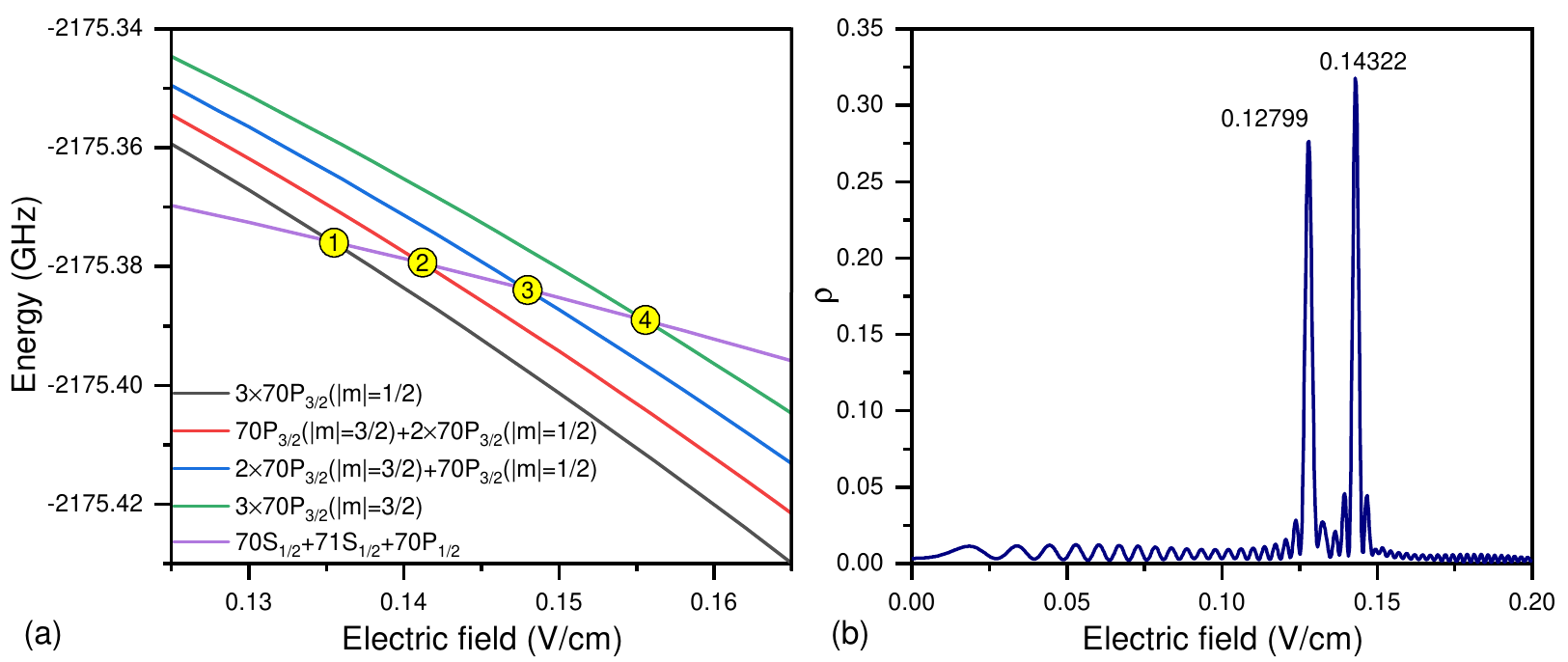}
\vspace{-.5cm}
\caption{
\label{Stark}
(a) Numerically calculated Stark structure of the collective energy levels, involved in three-body F\"{o}rster resonance $\ket{70P_{3/2}}^{\otimes 3} \to \ket{70S_{1/2};71S_{1/2};70P_{1/2}}$. The intersections 1-4 mark the positions of three-body resonances. (b) Numerically calculated dependence of the fraction $\rho$ of atoms in the final $\ket{71S_{1/2}}$ state after three-body interaction on the external dc electric field for the initial state $\ket{70P_{3/2}\left(m=1/2\right)}^{\otimes 3}$ (marked as 1 in Fig. 1(a)). The atoms are located along the Z axis at interatomic distance $R=10$ microns. 
}
\end{figure*}
\end{center}
%=========================================================================================================
%=========================================================================================================
%=========================================================================================================

\section{Introduction}

Recent advances in quantum information with ultracold neutral atoms include demonstration of quantum phases of matter on a large-scale quantum simulator~\cite{Ebadi2021}, demonstration of high-fidelity entanglement~\cite{Fu2021,Graham2019} and CNOT gates in atomic arrays with individual addressing~\cite{Levine2019}. These achievements are based on creation of defect-free arrays of optical dipole traps loaded with single atoms using spatial rearrangement by a movable optical tweezer~\cite{Schymik2020}, high-fidelity coherent laser Rydberg excitation of trapped atoms~\cite{Fu2021} and strong interatomic interaction, resulting in Rydberg blockade~\cite{Lukin2001, Saffman2010}. Note that the highest Bell state entanglement fidelities in atomic registers, recently demonstrated by Madjarov et al. are above 99.1\%~\cite{Madjarov2020}, which is close to the best results achieved on alternative platforms (ultracold ions~\cite{Srinivas2021}, superconductors~\cite{Yan2022}, etc.).

Of particular interest today is the implementation of controlled three-qubit quantum gates~\cite{Shi2018}, such as the Toffoli gate (described in Section~\ref{Sec3A}), the Deutsch gate, the Fredkin gate, etc. Such gates are key components for many important quantum algorithms, notably Shor's algorithm~\cite{Shor1999}, quantum error correction~\cite{Cory1998} and fault-tolerant computation~\cite{Dennis2001}. In addition, these gates greatly facilitate the implementation of quantum computing in large-scale registers.

Although multi-qubit gates can be decomposed into sequences of single-qubit and two-qubit gates, the lack of precision of two-qubit operations rapidly reduces the fidelity of composite multi-qubit gates~\cite{Morgado2021}. In this regard, we propose to use three-body F\"{o}rster resonances to implement three-qubit quantum operations.
F\"{o}rster resonance is a special type of dipole-dipole interaction that occurs when the energy levels of the collective states of atoms intersect in an external electric field~\cite{Pillet2009}. Depending on the number of atoms involved in the interaction, it can be either two-body~\cite{Ravets2014} or many-body~\cite{Faoro2015, Gurian2012}. A three-body F\"{o}rster resonance energy transfer (FRET) was first observed for an ensemble of $\sim 10^5$ cold Cs Rydberg atoms~\cite{Faoro2015}. This type of resonance corresponds to a transition when the three interacting atoms change their states simultaneously. One of the atoms here acts as a mediator of the interaction, carrying away the excess energy. This leads to a Borromean type of the F\"{o}rster energy transfer, when the ordinary two-body resonance gives a negligible contribution to the population transfer, as the three-body resonance appears at a different dc electric field, with respect to the two-body resonance. It thus represents an effective three-body operator, which can be used to directly implement Rydberg quantum gates.

Previously, we have demonstrated three-body F\"{o}rster resonances experimentally~\cite{Tretyakov2017} and proved theoretically that these resonances can be used to implement three-qubit quantum gates~\cite{Ryabtsev2018}. We have also designed a scheme of a high fidelity $\left(>98\%\right)$ three-qubit Toffoli gate for neutral atoms, based on the coherent phase dynamics of collective atomic states in the vicinity of such a resonance~\cite{Beterov2018a}. However, this scheme was quite complex for experimental implementation due to the need for individual excitation of the atoms into Rydberg states with different principal quantum numbers, as well as for extremely high precision electric field control and for adjustment of the positions of the resonances in electric field scale using an external magnetic field.

In this paper, we propose and theoretically investigate an improved scheme for the implementation of a three-qubit Toffoli quantum gate based on fine structure state changing three-body F\"{o}rster resonances, which we described in a previous paper~\cite{Cheinet2020}. In particular, we consider a scheme for implementing a fast, high-fidelity Toffoli quantum gate. A specific advantage of this scheme is the simplicity of its experimental implementation in large-scale registers (for interatomic distances of $\sim 10$ \textmu m). An analysis of the sensitivity of the gate fidelity to variations in experimental parameters is also given.

\section{Coherent three-body fine structure state changing F\"{o}rster resonances}

The dipole-dipole interaction operator between two neighboring atoms located along the quantization axis (Z) can be expressed as~\cite{Walker2008}:
\begin{eqnarray}
\label{eq1}
V_{dd} &=&\frac{e^{2} }{4\pi \varepsilon _{0} R^{3} } \left(\textbf{a}\cdot \textbf{b}-3a_{z} b_{z} \right)=\\
&=&-\frac{\sqrt{6} e^{2} }{4\pi \varepsilon _{0} R^{3} } \sum _{q=-1}^{1}C_{1q \; 1-q}^{20} a_{q} b_{-q}.  \nonumber
\end{eqnarray}
Here $\varepsilon_0$ is the vacuum dielectric constant; $e$ is the electron charge; \textbf{a} and \textbf{b} are the vectorial positions of the Rydberg electrons. The radial matrix elements of the dipole moment are calculated using a quasiclassical approximation~\cite{Kaulakys1995}.

As described in our article~\cite{Cheinet2020}, sets of three-body resonances are observed in real Rydberg atoms instead of a single resonance. This is due to the large number of interaction channels in real Rydberg systems. To reduce the number of observed resonances, it is necessary to choose the optimal geometry of the atomic register. We have proved that the linear arrangement of atoms at the same distances from each other along the quantization axis coinciding in the direction with the external control dc electric field (Z axis) is optimal~\cite{Ryabtsev2018}. Note that in this case the $V_{dd}$ operator couples only two-atom collective states with $\Delta M=0$, where \textit{M} is the total momentum projection of the collective state.  

Throughout the article, we will describe the behavior of the collective states of three Rb Rydberg atoms in the spatial configuration described above, as well as the interactions between them. These states have the form  $\ket{n_{1} l_{1} j_{1} \left(m_{j1}\right) ;n_{2} l_{2} j_{2} \left(m_{j2}\right) ;n_{3} l_{3} j_{3} \left(m_{j3}\right)} $ in Dirac notation. The F\"{o}rster energy defect is the difference between the energies of the final and initial collective states.

In our recent paper~\cite{Beterov2018a}, the following scheme of three-body F\"{o}rster resonance for ultracold Rb atoms was proposed for implementing a Toffoli gate:
\begin{eqnarray}
\label{eq2}
\ket{nP_{3/2};\left(n+1\right)P_{3/2};\left(n+1\right)P_{3/2} \left(m\right)} \to \\ \to \ket{nS_{1/2};\left(n+2\right)S_{1/2};\left(n+1\right)P_{3/2}\left(m^*\right)} \nonumber
\end{eqnarray}
Here $m$ is the projection of the angular momentum of the third atom on Z axis and $m^*$ is the changed value of this projection. It indicates that the states of all three atoms have been changed during the interaction. For $n=80$ our numeric simulations predicted relatively high fidelity ($\sim 98\%$) of the Toffoli gate operation. Nevertheless, the need to initialize atoms into states with different values of the principal quantum number is a significant difficulty for experimental implementation. Since the excitation of several atoms into different atomic states necessitates the simultaneous use of two or more lasers, the main problem during the experiment is the difficulty to keep the different radiation sources mutually coherent. Other limitations are related to the high sensitivity of the gate to changes in the electric field, as well as the need to use external magnetic field to adjust the positions of resonances. The alternative scheme of three-body resonances, proposed in~\cite{Cheinet2020}, is advantageous, since all atoms are initially excited into the same state:
\begin{eqnarray}
\label{eq3}
\ket{nP_{3/2}}^{\otimes 3} \to \ket{nS_{1/2};\left(n+1\right)S_{1/2};nP_{1/2}}
\end{eqnarray}
Note that $\ket{nP_{3/2}}^{\otimes 3}$ here denotes the product of three identical ket vectors. In this configuration of three-body interaction, the two-body F\"{o}rster resonance is known to be absent in rubidium for principal quantum numbers above $n=38$ due to the specific values of quantum defects and polarizabilities of Rydberg states $nP$ and $nS$ ~\cite{Cheinet2020,Walker2008}. Therefore, off-resonant two-body interactions induce small phase shifts but no sizable population transfer, in contrast to the scheme, previously considered in~\cite{Beterov2018a}. This substantially simplifies the population and phase dynamics of the collective three-body states, as will be shown below.

In Figure~\ref{Stark}(a), the energies of the collective Rydberg states involved in three-body F\"{o}rster resonance (\ref{eq3}) are depicted as functions of the external electric field. These dependencies of energy levels are calculated for different fine structure components of Rb $70P$ state. The intersections with the final quantum state, indicated as 1-4, mark the positions of possible three-body resonances.

Figure~\ref{Stark}(b) shows the dependence of the calculated probabilities of the F\"{o}rster resonant energy transfer on the external electric field when all atoms are initially in the state $\ket{70P_{3/2}\left(m=1/2\right)}$. This corresponds to case~1 in Fig.~\ref{Stark}(a). Two resonant features are clearly seen. The splitting and shift of the resonances are caused by multiple channels of three-body F\"{o}rster interaction through different intermediate quantum states.

%=========================================================================================================
%=========================================================================================================
%=========================================================================================================
\begin{center}
\begin{figure}[!ht]
\center
\includegraphics[width=\columnwidth]{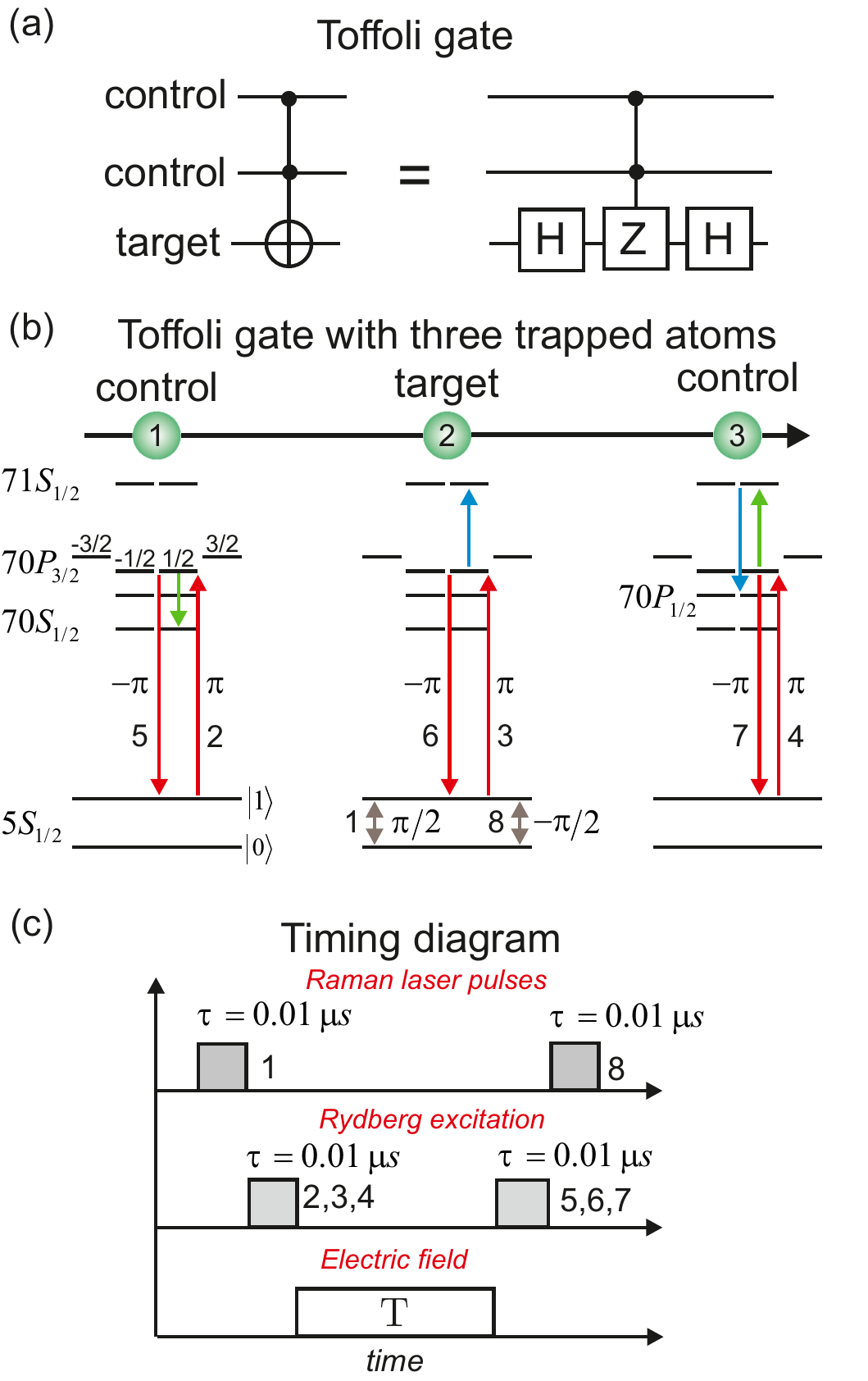}
\vspace{-.5cm}
\caption{
\label{Scheme}
(a) General scheme of the three-qubit Toffoli gate. (b) Scheme of the Toffoli gate based on three-body Rydberg interactions. Three atoms are located in the individual optical dipole traps aligned along the Z axis, which is co-directed with the controlling dc electric field. Laser Raman (or microwave) pulses 1 and 8 drive transitions between the logical states $\ket{0}$ and $\ket{1}$ of the target qubit. Laser pulses 2-7 excite and de-excite the chosen Rydberg states of the three atoms. The $\pi$ phase shift due to the three-body interaction appears only if all three atoms are excited into Rydberg states. The green and blue arrows here indicate $\ket{70P_{3/2}}^{\otimes 3} \to \ket{70S_{1/2};70P_{3/2};71S_{1/2}}$ and $\ket{70S_{1/2};70P_{3/2};71S_{1/2}} \to \ket{70S_{1/2};71S_{1/2};70P_{1/2}}$ intermediate two-body transitions, respectively. (c) Timing diagram of the pulses in the proposed gate scheme. The whole gate scheme includes the following 5 steps: application of pulse 1, simultaneous application of pulses 2-4, application of a constant external electric field, simultaneous application of pulses 5-7, application of pulse 8.
}

\end{figure}
\end{center}

%=========================================================================================================
%=========================================================================================================
%=========================================================================================================

%=========================================================================================================
%=========================================================================================================
%=========================================================================================================
\begin{center}
\begin{figure*}[!htb]
\center
\includegraphics[width=\textwidth]{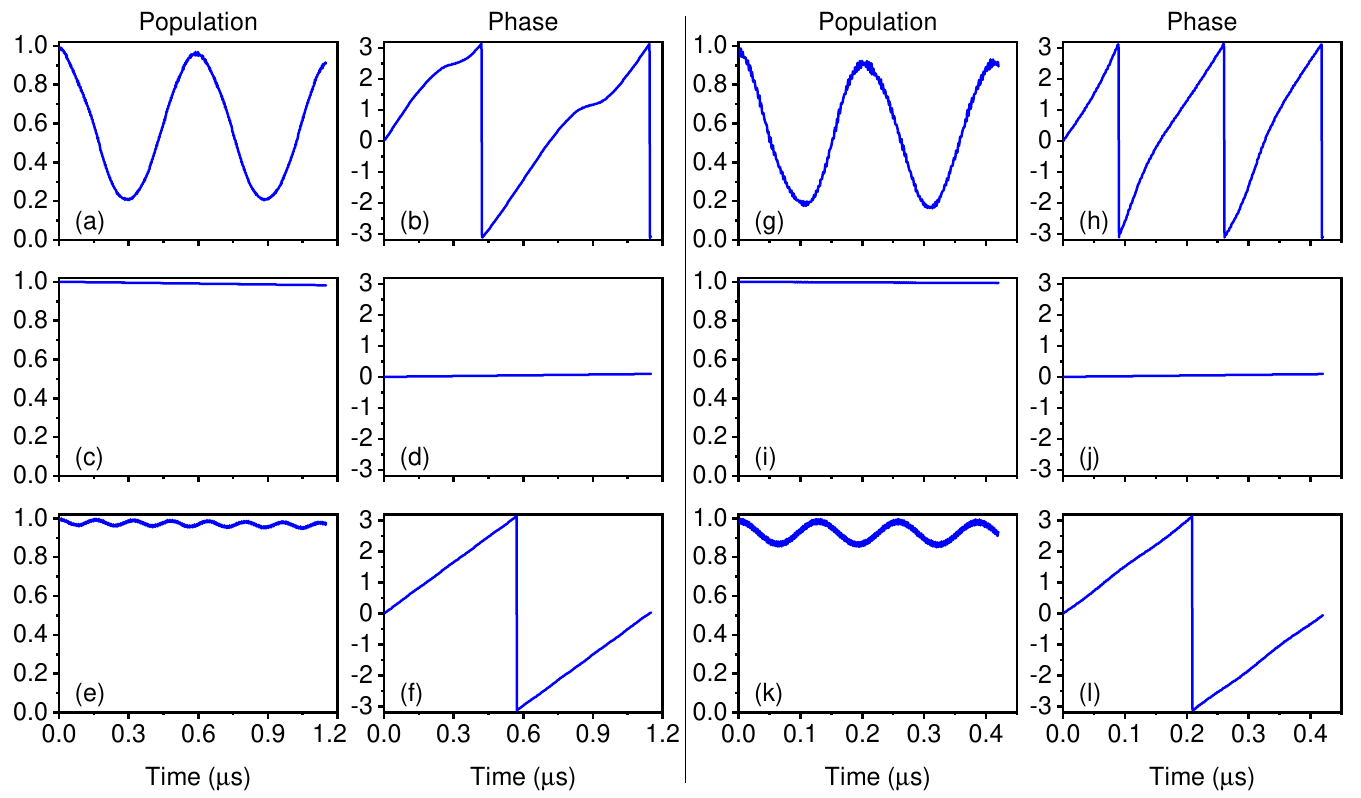}
\vspace{-.5cm}
\caption{
\label{Phase_diag}
Numerically calculated time dependences of the populations and phases of the initially excited collective states of three interacting atoms. The upper row (a, b, g, h) depicts the multiparticle state population and phase evolution when all three atoms are excited into Rydberg states ($\ket{rrr}$). The middle (c, d, i, j) and lower (e, f, k, l) rows belong to configurations $\ket{rgr}$ and $\ket{grr}$ ($\ket{rrg}$), respectively. Here $\ket{g}$ is the ground state which can be either $\ket{0}$ or $\ket{1}$, $\ket{r}$ is the Rydberg state $\ket{70P_{3/2}\left(m=1/2\right)}$ The phase values are presented in ordinary units in the range ($-\pi$, $\pi$). System parameters: (a - f) $R = 10$ \textmu m; $E=0.14235$ V/cm; $T = 1.15$ \textmu s; (g - l) $R = 8.5$ \textmu m;$E=0.1469$ V/cm; $T= 0.42$ \textmu s.
}
\end{figure*}
\end{center}
%=========================================================================================================
%=========================================================================================================
%=========================================================================================================

\section{The Toffoli gate}

\subsection{The gate scheme}
\label{Sec3A}

The Toffoli quantum gate (or CCNOT gate) is a universal three-qubit quantum gate. It is very important for the effective implementation of many quantum algorithms, in particular, for quantum error correction. This gate can also be represented as a CCZ gate wrapped with Hadamard gates, as shown in Fig.~\ref{Scheme}(a).

The implementation of the Toffoli quantum gate in a system of neutral atoms was described by Levine et al. in~\cite{Levine2019}. The proposed implementation is based on a strong blockade of the nearest neighbors in a trimerized 1D array. The achievable gate fidelity in this case was $F=0.87(4)$(after state preparation and measurement (SPAM) errors correction). These results compare quite well with Toffoli gate implementations with trapped ions ($ F=0.896$~\cite{Figgatt2017}) and superconducting circuits ($F=0.78$  ~\cite{Reed2012}). However, these values are far from the threshold required for the implementation of fault-tolerant quantum computing in atomic registers ($F\geq 0.99$). Equally important is the fact that quantum gates based on the Rydberg blockade effect require a sufficiently close arrangement of atoms~\cite{Isenhower2011}. The use of strong resonant dipole-dipole interactions is one of the promising solutions for working in large-scale registers, where it is required to implement gates between qubits spatially isolated from each other at distances of about $10$ microns or more. In this case, possible quantum gate schemes can be based on adiabatic passage along the dark state of the Rydberg system~\cite{Khazali2020}, as well as on the use of F\"{o}rster resonances, which are studied in this article.

The proposed scheme for the implementation of the Toffoli gate is shown in Fig.~\ref{Scheme}(b). Three Rb atoms are confined in three optical dipole traps located along the direction of the external electric field (Z axis) with interatomic distance R.
To couple the logical states of qubits (namely, $\ket{0}$ and $\ket{1}$), we propose to use two-photon Raman laser pulses that do not populate the intermediate excited state $5P$. An alternative approach based on the use of microwave laser pulses with individual addressing can also be applied. This will require the use of an intense off-resonant laser acting on a selected qubit to ac Stark shift its energy levels into resonance with the microwave radiation~\cite{Weitenberg2011,Zhang2006,Xia2015}. A three-photon excitation scheme can be used to pair the logical states of qubits with Rydberg levels~\cite{Ryabtsev2011}. The effects associated with the phase and intensity noise of the laser were considered in detail by De L\'{e}s\'{e}leuc et al. in~\cite{De2018}. 

Eight laser pulses are used to implement the gate. As the first step, the pulse 1 is used, which is a $Y$-rotation by $\pi/2$, carrying out the action of the first Hadamard gate. Then, the pulses 2-4 required for the $\ket{1} \to \ket{70P_{3/2}\left(1/2\right)}$ transitions are applied simultaneously to all three qubits. The number in parentheses indicates the projection of the momentum $m_j$ on the Z axis. 

In accordance with the proposal~\cite{Cheinet2020}, we consider the excitation into Rydberg states with the same principal quantum number $n=70$. This configuration allows us to achieve high fidelity of the quantum gate due to long lifetimes, large dipole moments and coherence of the chosen three-body interaction channels. At the same time, it facilitates the experimental implementation of the scheme, compared to our previous proposal ~\cite{Beterov2018a}.

Depending on the initial state of the system, after laser pulses 2-4 have been applied, the number of the excited Rydberg atoms varies from zero to three. When all three atoms are excited, the phase of the collective atomic state is shifted by $\pi$, due to the three-body F\"{o}rster resonance, tuned by an external electric field.

At the final stage, Rydberg atoms are de-excited by laser pulses 5-7. Raman laser or microwave pulse 8 drives the additional $-\pi /2$ rotation of the target qubit around the Y axis, which is equivalent to the second Hadamard gate in Fig.~\ref{Scheme}(a). The timing diagram of all controlling pulses is shown in Fig.~\ref{Scheme}(c).

To calculate the phase and population dynamics of the atomic system, we used the method described in~\cite{Beterov2018a}. We solved the non-Hermitian Hamiltonian based Schr\"{o}dinger equation for the probability amplitudes of the 360 collective states taking into account Rydberg lifetimes~\cite{Beterov2009}. For simplicity, we considered an open system and neglected the return of the population from Rydberg states to the ground states due to spontaneous decay.

\subsection{Phase and population dynamics}

To implement the Toffoli gate, it is necessary to find the conditions under which different interatomic interactions lead to the required phase shifts of the initially excited collective states. Therefore, it is necessary to optimize the parameters of the atomic system: the interatomic distance $R$, the interaction time $T$ and the value of the external dc electric field. Taking into account the technical limitations of experimental implementations, it is necessary to pay attention to the required accuracy of the parameter values. In particular, we found the following requirements for accuracy thresholds: the interatomic distance must be controlled with an accuracy of $0.1$ \textmu m; the interaction time~-~$0.01$ \textmu s; the external electric field~-~$10^{-4}$ V/cm. Here we assume that the maximum allowable deviation of the gate fidelity cannot exceed one percent.

Figure~\ref{Phase_diag} shows the numerically calculated phase and population dynamics of the initially excited collective two- and three-body Rydberg atomic states for optimized system parameters. Left-hand and right-hand panels of Fig.~\ref{Phase_diag} show the time dependencies of the populations and phases of initial collective states for interatomic distances $R=10$ \textmu m and $R=8.5$ \textmu m, respectively. When calculating the gate scheme, both two-body and three-body interactions in the atomic system were taken into account. Note that for the successful execution of the gate, it is extremely important that the populations of the initial states are close to unity after the end of the interaction.

If all three atoms are excited into Rydberg states, we observe almost resonant Rabi-like population oscillations (Figs.~\ref{Phase_diag}(a, g)). In this case, the phase of the state changes by $\pi$ after the interaction time due to the three-body resonance $\ket{70P_{3/2}\left(m=1/2\right)}^{\otimes 3} \to \ket{70S_{1/2};71S_{1/2};70P_{1/2}}$ (Fig.~\ref{Phase_diag}(b, h)). This phase shift is sensitive to the electric field which acts directly on the F\"{o}rster energy defect. It corresponds to the controlled phase shift when all three atoms are in state $\ket{1}$ prior to the Rydberg excitation in Fig.~\ref{Scheme}(b). Note that Fig.~\ref{Scheme}(b) shows only one of the possible transition schemes. In the resonant process, we cannot attribute $\ket{70P_{3/2}} \to \ket{70S_{1/2}}$, $\ket{70P_{3/2}} \to \ket{71S_{1/2}}$ and $\ket{70P_{3/2}} \to \ket{70P_{1/2}}$ transitions to a specific atom 1, 2 or 3. The population of initial state after the completion of the interaction is $91.5\%$ due to the finite Rydberg lifetimes and the leakage of population to other collective levels by Rydberg interactions. These are found to be the major sources of the gate error.

Consider the case when only two of the three atoms are in Rydberg states. Then, due to the selection rule $\Delta M=0$ only off-resonant two-body  interactions $\ket{70P_{3/2}\left(m=1/2\right)}^{\otimes 2}\leftrightarrow \ket{70S_{1/2}\left(m=1/2\right);71S_{1/2}\left(m=1/2\right)}$ are possible. The state $\ket{70S_{1/2};71S_{1/2}}$ can also interact off-resonantly with $\ket{70P_{1/2}}^{\otimes 2}$  and $\ket{70P_{3/2};70P_{1/2}}$  states. A detailed analysis of two-body interactions in three-body systems of Rydberg atoms is given in our other article~\cite{Pham2020}.

If the atomic ensemble is initially excited to state $\ket{rgr}$ (here $\ket{g}$ is the ground state which can be either $\ket{0}$ or $\ket{1}$, $\ket{r}$ is the Rydberg state $\ket{70P_{3/2}\left(m=1/2\right)}$), we can consider the influence of the interactions described above on the population and the phase of the final state as negligible (Figs.~\ref{Phase_diag}(c, d, i, j)). This is due to the fact that two Rydberg atoms are too far apart from each other. According to equation~(\ref{eq1}), an increase in the distance between atoms by a factor of two causes an eightfold decrease in the strength of the dipole-dipole interaction.

Alternatively,when the ensemble is initially excited into one of the states $\ket{grr}$ or $\ket{rrg}$, we can observe a significant influence of the off-resonant two-body interactions on the phase of the collective state (Figs.~\ref{Phase_diag}(f, l)). This leads to the phase shift of the initially excited state, which can be compensated to zero during the interaction time $T$ (see Fig.~\ref{Scheme}(c)). This phase shift is found to be sensitive to the external electric field. Two-body interactions also have a significant impact on the population of the initial state, leading to weak (with an amplitude of$~5-10\%$) Rabi-like oscillations (Figs.~\ref{Phase_diag}(e, k)).

Finally, when only one atom in the ensemble is temporarily excited into the Rydberg state, the $\pi$ and $-\pi$ pulses, shown in Fig.~\ref{Scheme}(b), will return the system into the initial state with zero phase shift. However, temporary Rydberg excitation will result in population loss due to the finite lifetimes of Rydberg states. The trivial case is when no Rydberg atoms are excited. The pulses 2-7 will have no effect in this instance.

In contrast to our previous proposal~\cite{Beterov2018a}, we obtained the required phase dynamics without the need to use an external magnetic field for fine tuning of the position of three-body F\"{o}rster resonance in the electric field scale. Moreover, the absence of the two-body F\"{o}rster resonance in the vicinity of the three-body F\"{o}rster resonance substantially simplifies the phase dynamics of the collective three-atom states.

\subsection{Optimization of gate parameters}

%=========================================================================================================
%=========================================================================================================
%=========================================================================================================
\begin{center}
\begin{figure}[!h]
\center
\includegraphics[width=\columnwidth]{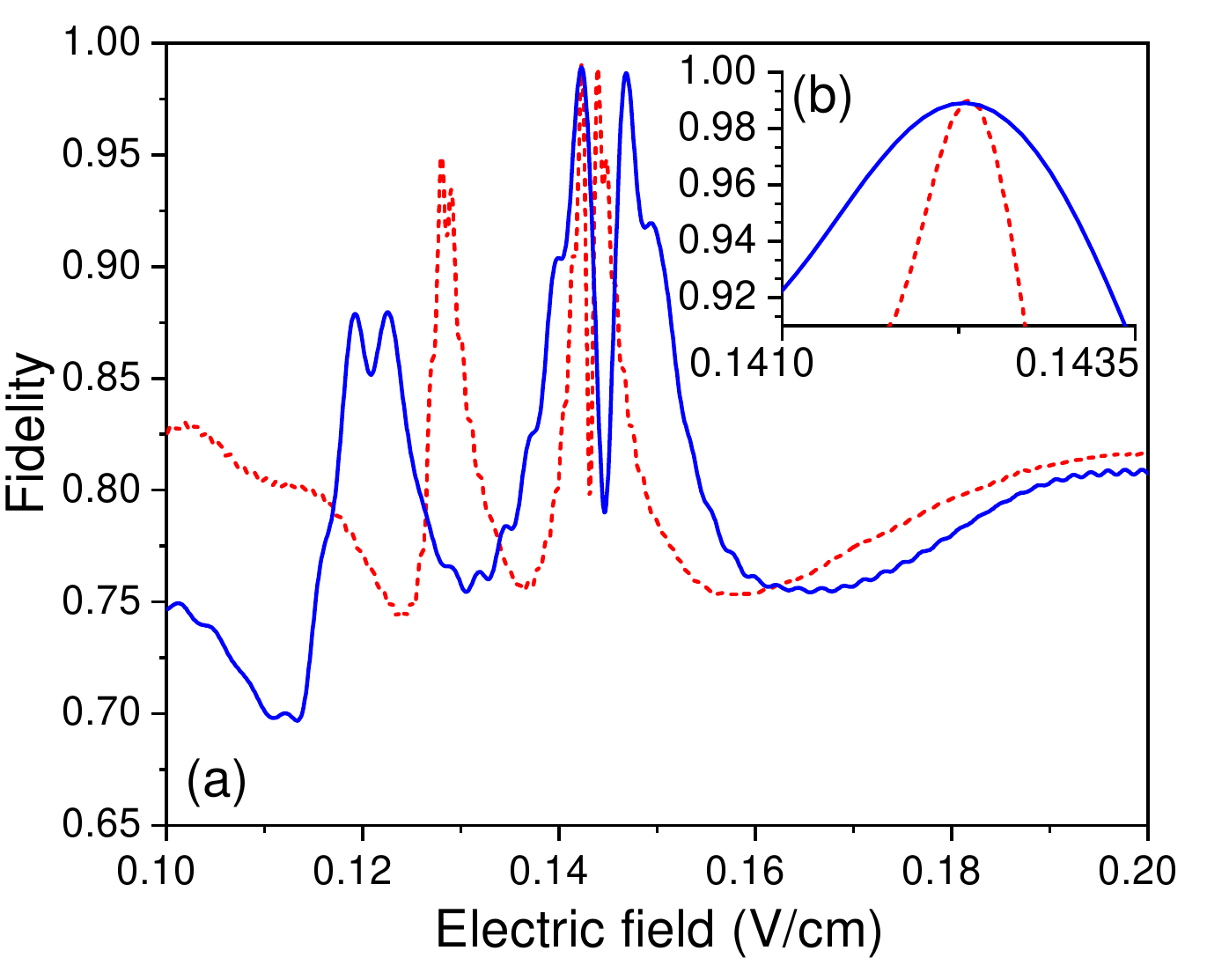}
\vspace{-.5cm}
\caption{
\label{Fidelity}
Dependence of the fidelity of the Toffoli gate on the dc electric field for two different interatomic distances: $R=10$ \textmu m (red curve) and $R=8.5$ \textmu m (blue curve). The maximum fidelity of $99.05\%$ is achieved with an electric field of $0.14232$ V/cm. The interaction times coincide with those indicated in the description of Fig.~\ref{Phase_diag} for both cases.(a) For a wide range of electric field values (0.1-0.2 V/cm). (b) Near the fidelity maxima.
}
\end{figure}
\end{center}

%=========================================================================================================
%=========================================================================================================
%=========================================================================================================

The optimal values of the system parameters (interatomic distance R, interaction time T, electric field value E) were calculated by performing multi-objective optimization using the Nelder-Mead method in order to increase the gate fidelity. As mentioned above, for experimental implementation, these parameters must be controlled with high accuracy. Thus, when developing a gate scheme, it is necessary to take into account all possible sources of the gate fidelity losses arising from insufficient control of parameters and suggest ways to minimize their total effect.

The greatest control accuracy is necessary for the dc electric field: as can be seen from Fig. 1, the resonance peaks are extremely narrow, and even a field variation of $10^{-4}$ V/cm can critically affect the gate fidelity. To mitigate this disadvantage, we propose to reduce the interatomic distances.

Figure~\ref{Fidelity} shows the dependence of the gate fidelity on the external electric field for two different interatomic distances. It can be seen that with a decrease in distance the requirements for the accuracy of field value control are significantly reduced. At a distance of $R=10$ \textmu m, a fidelity loss of $1\%$ (with a maximum fidelity of $99.05\%$) is obtained for a field mismatch of $10^{-4}$ V/cm. At $R=8.5$ \textmu m, the same fidelity loss is obtained only at a field mismatch of about $4\cdot 10^{-4}$ V/cm.

It should also be noted that the distance reduction has a  positive effect on the timing of the quantum gate.
Specifically, the time required for gate implementation is 0.42 \textmu s when the distance between atoms is 8.5 \textmu m. In the case when the interatomic distance is 10 microns, the required time is about 3 times higher.

To estimate the gate fidelity, the method proposed in~\cite{Bowdrey2002} was used. We considered 6 single-qubit configuration states: $\ket{0}$, $\ket{1}$, $\left( \ket{0}+\ket{1} \right)/ \sqrt{2}$, $\left( \ket{0}-\ket{1} \right)/ \sqrt{2}$, $\left( \ket{0}+i\ket{1} \right)/ \sqrt{2}$ and $\left( \ket{0}-i\ket{1} \right)/ \sqrt{2}$. We formed a set of three-qubit states as all $6^3=216$ combinations of three single-qubit basis states. We simulated the density matrices $\rho_{sim}$ of all final states after Toffoli gate was applied to each initial state. Then we calculated the fidelity of each final state comparing to the etalon state $\rho_{et}$, which is the final state of the ensemble after the perfect Toffoli gate is performed~\cite{Nielsen2011}:
\begin{eqnarray}
\label{eq5}
F=\textrm{Tr}\sqrt{\sqrt{\rho_{et}}\rho_{sim}\sqrt{\rho_{et}}}
\end{eqnarray}

Averaging over all 216 states, we calculated the gate fidelity of $99.05\%$.

Note that the losses of the gate fidelity occurring at the stages of excitation and de-excitation of the Rydberg levels were not taken into account in this calculation. These losses arise mainly due to degeneracy of the Zeeman sublevels of Rydberg atoms in a zero electric field, which leads to undesirable two- and three-body interactions between the collective Rydberg states. According to our estimate, the maximum fidelity leakage caused by these processes does not exceed 0.88\%. To reduce this effect, it is possible to conduct excitation in an external electric field. We found that using the electric field of 0.2 V/cm one can reduce the fidelity loss to 0.23\%. Additional multiparametric optimization allows one to adjust the fidelity values to fully compensate for the described effect. Thus, a fidelity loss of 0.03\% was obtained for the Toffoli gate model with the following parameter values: $R = 10$ \textmu m; $E=0.14225$ V/cm; $T = 1.12$ \textmu s, $E_0=0.2$ V/cm, $\tau = 0.01$ \textmu s. Here $E_0$ is the described external excitation field, and $\tau$ is the duration of the excitation pulse. In this study, these two parameters were chosen for analytical reasons and were not included in the optimization process.
Finally, we can summarize that the theoretical fidelity of the proposed gate is $F>99\%$.

Since the excitation electric field value and the durations of the exciting and de-exciting pulses are variable parameters, they provide an additional opportunity to control the interaction in a three-body system. We are confident that by performing multiparametric optimization taking into account these parameters, the gate fidelity can be significantly increased. A similar approach to boost the fidelity of quantum operations was demonstrated in article~\cite{Delvecchio2022}. However, this issue requires additional research.

As mentioned above, the limited lifetimes of Rydberg states are major sources of the gate error in the proposed gate scheme. A possible solution to this problem may be the use of a cryogenic environment~\cite{Mack2015}.

%=========================================================================================================
%=========================================================================================================
%=========================================================================================================

\section{Conclusion}

In this paper we proposed a scheme to implement a three-qubit Toffoli gate based on a new type of three-body resonant F\"{o}rster energy transfer in the ensemble of Rb Rydberg atoms isolated in three individual optical dipole traps. This new type of resonance is based on a change of fine structure state of one of the atoms involved in the interaction. The collective phase shifts induced by Rydberg interactions are controlled by an external electric field. We have shown that it is possible to reach a fidelity exceeding $99\%$ for a short gate duration from 0.4 \textmu s to 1.2 \textmu s.

Note that in the proposed scheme, off-resonant two-body interactions lead to relatively weak phase dynamics. This reduces the effect of the complex structure of Rydberg energy levels on gate fidelity, which appears to be the major source of gate error if the Rydberg interactions are strong~\cite{Derevianko2015}. This also makes it possible to implement quantum gates in large-scale registers (for interatomic distances of $\sim 10$ \textmu m).

Compared to our previous proposal~\cite{Beterov2018a}, the improved scheme of Toffoli gate is more suitable for experimental implementation, since the initial states of atomic qubits are completely identical. It also does not require the use of a magnetic field to fine-tune the resonance position. In order to minimize the decrease in gate fidelity, we found a compromise between the control accuracies of various experimental parameters (interaction time, interatomic distance and dc electric field value). We managed to achieve a significant reduction in the sensitivity of the circuit to electric field deviations by reducing the interatomic distance, avoiding a decrease in the gate fidelity.

%=========================================================================================================
%=========================================================================================================
%=========================================================================================================

\begin{acknowledgments}

This work was supported by the Russia-France cooperation grant ECOMBI (CNRS grant No. PRC2312 and RFBR grant No. 19-52-15010). The Russian team was also supported by the Novosibirsk State University and by the Foundation for the Advancement of Theoretical Physics and Mathematics \enquote{BASIS}. The French team was also supported by the EU H2020 FET Proactive project RySQ (grant No. 640378).

\end{acknowledgments}

%\bibliography{JCbib}{}
%apsrev4-2.bst 2019-01-14 (MD) hand-edited version of apsrev4-1.bst
%Control: key (0)
%Control: author (72) initials jnrlst
%Control: editor formatted (1) identically to author
%Control: production of article title (-1) disabled
%Control: page (0) single
%Control: year (1) truncated
%Control: production of eprint (0) enabled
%

\end{document}